\documentclass[twocolumn]{aastex62}
\usepackage{natbib,twoopt,amsmath}
\usepackage{color}
\hypersetup{linkcolor=red,citecolor=green,filecolor=cyan,urlcolor=magenta}
\received{Jan 3, 2019}
\accepted{Jan 24, 2019}
\submitjournal{ApJ}

\shorttitle{Cusp-type magnetic null point}
\shortauthors{Nickeler et al.}
\begin{document}
\title{Topological structures of velocity and electric field in the vicinity of a cusp-type magnetic null point}

\correspondingauthor{Dieter H. Nickeler}
\email{dieter.nickeler@asu.cas.cz}

\author[0000-0001-5165-6331]{Dieter H. Nickeler}
\affiliation{Astronomick\'y \'ustav, Akademie v\v{e}d \v{C}esk\'e republiky, v.v.i., Fri\v{c}ova 298, 251\,65 Ond\v{r}ejov, Czech Republic}

\author[0000-0002-3963-8701]{Marian Karlick\'y}
\affiliation{Astronomick\'y \'ustav, Akademie v\v{e}d \v{C}esk\'e republiky, v.v.i., Fri\v{c}ova 298, 251\,65 Ond\v{r}ejov, Czech Republic}

\author[0000-0002-4502-6330]{Michaela Kraus}
\affiliation{Astronomick\'y \'ustav, Akademie v\v{e}d \v{C}esk\'e republiky, v.v.i., Fri\v{c}ova 298, 251\,65 Ond\v{r}ejov, Czech Republic}

\begin{abstract}
Topological characteristics reveal important physical properties of plasma structures and
astrophysical processes. Physical parameters and constraints are linked with topological invariants, 
which are important for describing magnetic reconnection scenarios. We analyze stationary 
non-ideal Ohm's law concerning the Poincar\'{e} classes of all involved physical fields in 2D by 
calculating the corresponding topological invariants of their Jacobian (here: particularly 
the eigenvalues) or Hessian matrices. The magnetic field is assumed to have a cusp 
structure, and the stagnation point of the plasma flow coincides with the cusp. We find that the 
stagnation point must be hyperbolic. Furthermore, the functions 
describing both the resistivity and the Ohmic heating have a saddle point structure, being displaced
with respect to the cusp point. These results imply that there is no monotonous relation between current
density and anomalous resistivity in the case of a 2D standard magnetic cusp.
\end{abstract}


\section{Introduction}

Magnetic reconnection is a key process for understanding magnetic structures and their topological change 
in astrophysical plasmas. Many investigations analyzing especially topological or geometrical properties 
of physical fields, focusing either on some or involving all of them, have been done in various contexts 
and approaches: for example resistive magnetohydrodynamics (MHD) in 2 dimensions (2D) 
\citep{1975JPlPh..14..271P}, kinematic ideal MHD in 3D \citep{1990ApJ...350..672L}, purely 
magnetically in 2D and 3D \citep{1996PhPl....3..759P}, resistive MHD in 3D  with constant resistivity
\citep{2000PhPl....7.3542T}, resistive MHD in 3D with locally varying but prescribed resistivity 
\citep{2003PhPl...10.2712H, 2009PhPl...16l2101P}, Hall-MHD in 2.5D \citep{2009ApJ...694.1464L}, 
resistive, kinematic MHD in 3D with varying resistivity \citep{2011JPlPh..77..843W}, and resistive MHD 
in 2D with localized but consistent resistivity \citep{2012AnGeo..30.1015N}.

The origin of these investigations and definitions of geometrical shapes of field lines in the vicinity
of singular points of vector fields can be attributed to \citet{hp1881jm} and \citet{Lyapunov}. 
To describe vector fields and functions around null points qualitatively, it is inevitable to 
identify topologically equivalent structures of vector fields or scalar fields, i.e., phase portraits of
corresponding dynamical systems. Fields with topologically equivalent structures were gathered by 
\citet{hp1881jm} in classes, to which we refer to as Poincar\'{e} classes. The classification is based 
on invariants of the Jacobian matrices (i.e. their eigenvalues) of these vector fields. In analogy, 
scalar functions, such as the resistivity, can also be classified locally by invariants related to 
their Jacobian and Hessian matrices. The eigenvalues hence serve as sort of a fingerprint of the 
specific shape of the vector field. The analysis may result in both structurally stable as well as 
unstable fields, whereby in the latter case the solutions might be degenerated.

Different magnetic and flow null points exist defining the various classes. For instance, in linearized 
systems hyperbolic null points like X-type null points in 2D (Poincar\'{e} class of hyperbolic or saddle 
points), or elliptic null points in 2D (Poincar\'{e} class of elliptic or O-points) can 
exist\footnote{For a comprehensive overview of these and other classes see \citet{amann} Chapter III, 
Section 13 or \citet{arnold_ODE} Chapter 3.}. Explicitly, the flow around an obstacle at a 
stagnation point is of hyperbolic or X-point 
type while both eigenvalues are real, whereas two purely imaginary eigenvalues with absolute value of  
one describe a so-called center where the stream lines are topological circles surrounding an O-point.
We would like to stress that in linearized systems for magnetic fields or, more generally speaking,
divergence free fields such as incompressible velocity fields, only X- and O-points can exist in 2D.
However, also other types of null points of higher order can occur, such as degenerated null points 
like cusps \citep[e.g.,][]{arnold}. 

The mathematical tools to analyze differential topological properties concerning the nature of 
dynamical systems are provided by the algebraic structure of functions:
Fields that are not linear (first derivative, Jacobian matrix) in the vicinity of a null point 
require a representation with polynomials of at least second order. This means that higher derivatives 
(Hessian matrix) start to play an important role concerning their invariant properties.

Imprinting arbitrary properties on a vector field may lead to a loss of information of structural
characteristics of this field. For example, constraints in the frame of MHD or restrictions of the 
(static or stationary) solutions, such as demanding the existence of singular current (or vortex) 
sheets, constant resistivities, etc., can lead to structural unstable solutions for one or more of the 
physical fields. This destroys the generic property of the vector fields or functions, describing 
physical parameters, and hampers their unambiguous classification.

The topological and geometrical structures of the different field lines (e.g. electric, velocity,
current, magnetic field) as a result of magnetic reconnection or similar processes
where singular points of the magnetic field are involved, can be determined 
by identifying the Poincar\'{e} classes of the various physical fields.
The Poincar\'{e} classes of flows and total electric field in the frame of non-ideal MHD are not 
independent, if a certain Poincar\'{e} class of a magnetic null point is present. 
How the plasma flow passes through the null point domain depends on the Poincar\'{e} class of the total 
electric field and restricts which Poincar\'{e} classes the flow can take.

The analysis of topological properties is not only a mathematical exercise, but it reveals 
dissipation processes and sheds light on many astrophysical problems concerning plasma dynamics
such as the heating of the solar corona \citep[e.g.,][]{1972ApJ...174..499P, 1994ISAA....1.....P, 
2010ApJ...718..717L, 2015ApJ...808..134C}.

In 2D, the search of the Poincar\'{e} class of the total electric field is linked to the Poincar\'{e} 
class of the resistivity. It is well known that for exact and analytical reconnection solutions, it is 
not sufficient to assume any non-ideal term or non-idealness or for example a constant resistivity 
\citep{1994JGR....9921467P, 1995ApJ...450..280C, 1996PhPl....3.3188N, 1998ApJ...505..363W, 2000JPlPh..64..601T, 2000PhPl....7.3542T,
2012AnGeo..30.1015N, 2014A&A...569A..44N}. For the classical role of reconnection in 2D it is inevitable that the plasma flow 
can cross {\it both} magnetic separatrix branches, as this scenario constitutes the 
reconnection solution. This characteristic does often not exist if a constant resistivity is assumed 
\citep{1995ApJ...450..280C}, or if the non-idealness shows a one-dimensional character, like in the case 
of current (sheet) depending resistivities. Very often such resistive or non-ideal terms only allow for 
crossing one of the two branches of the separatrix \citep{2012AnGeo..30.1015N}, which are the so-called reconnective annihilation
solutions. 

In the general case, the resistivity is not necessarily only depending on state variables, but also on 
higher derivatives of the state variables and on the topological and geometrical electromagnetic field structure. 
Therefore, to generate a non-linear perturbation on the right hand side of ideal Ohms's law, it is not 
necessarily sufficient to consider only two- or multi-fluid effects or a kinetic ansatz, because 
deviations from ideal MHD can also be caused by other physical processes \citep[see, 
e.g.,][]{1984ApJ...276..391S}.

If the magnetic skeleton is given, for example by prescribing the Poincar\'{e} class of its null points, 
then the question arises, which types of skeleton of the other physical fields (total electric field 
respective resistivity, 
velocity field) are allowed. 
The natural way of determining the geometry is to assume that the fields
are locally regular and do not contain singular current sheets, so, for example, the real cusps must be 
finite and surrounded by smooth field lines. 

 \begin{figure}[t]
   \centering
    \includegraphics[width=\hsize]{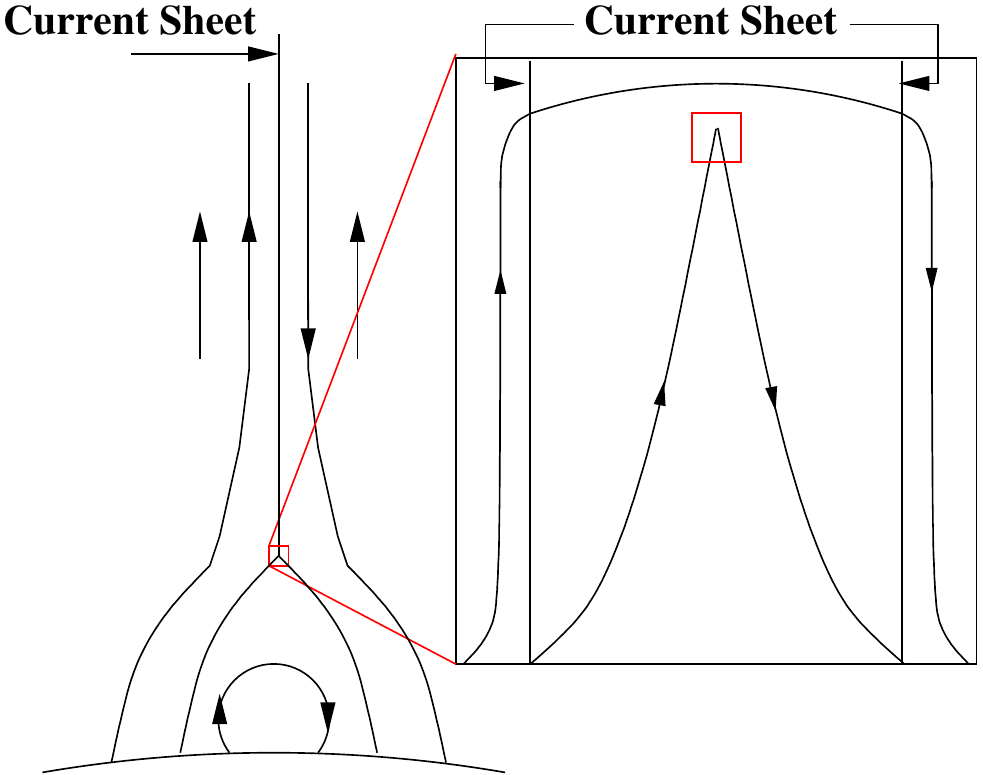}
      \caption{Sketch of a helmet streamer with cusp point and singular current sheet and zoom to the inner cusp point region within the current sheet. The tiny red box inside the zoom defines the domain of our analysis.
      }
         \label{fig:helmet}
  \end{figure}
  
Observations reveal magnetic structures in the solar atmosphere such as coronal streamers or helmet 
streamers that display a cusp-like structure \citep{1992SSRv...61..393K, 2007ApJ...671..912U} with a 
postulated current sheet having an extremely small width compared to the typical scale-length of the
total structure. This small width of the current sheet is emulated in theoretical models by a singular 
current sheet (left part of Figure~\ref{fig:helmet}), implying that the models focus on large scales 
rather than on the small-scale field structures allowing for a Poincar\'{e} classification.  In the 
scenario of a singular current sheet the transverse component of the field at the cusp with respect
to the current sheet axis is zero. The analysis of such non-regular cusp structures 
in the literature were performed either analytically \citep[e.g.,][]{1992ApJ...384..333V, 
1993SoPh..146..119V, 1997PhPl....4.3960U, 1998SoPh..180..439W, 2000SoPh..191..391W} or based on 
numerical simulations \citep[e.g.,][]{2007AdSpR..39.1415K}. In all these investigations the 
inner structure of the current distribution was neglected.

In this work, we investigate the internal structure of the current sheet at magnetic cusps. We 
concentrate on regular, mathematical cusp structures with a finite scale of the current 
distribution, in contrast to the singular current sheets mentioned before. This means that we are 
considering regular magnetic flux functions in the very close vicinity of the singular magnetic point, 
here the cusp point, to characterize the topological connections of the involved physical fields and to 
allow for transverse components of the magnetic field inside the current sheet (right part of 
Figure~\ref{fig:helmet}).

\section{Objectives and methodology}

The investigation of topological properties not only of the magnetic field $\textbf{\textit{B}}$, but 
also of the plasma flow $\textbf{\textit{v}}$ and the non-idealness $\textbf{\textit{N}}$, requires the 
violation of ideal Ohms's law, $\textbf{\textit{E}}+\textbf{\textit{v}}\times \textbf{\textit{B}} = 
\textbf{0}$. The reason is that in 2D the electric field is $\textbf{\textit{E}} = E_{z}
\textbf{\textit{e}}_{z}$, where $E_{z} = \rm{const}$ in the case of stationary translational invariant 
MHD. If the electric field component in $z$-direction does not vanish, i.e. $E_{z} \neq 0$, no 
regular solutions exist for the velocity field at a magnetic null point, meaning that the velocity field 
diverges. Therefore, one typically has to use non-ideal Ohms's law,  
\begin{equation}
\textbf{\textit{E}}+\textbf{\textit{v}}\times\textbf{\textit{B}} = \textbf{\textit{N}}\, ,
\label{ohm1}
\end{equation}
to describe flows around a magnetic null point. In case of resistive MHD, the 
non-ideal term (or total electric field) $\textbf{\textit{N}}$ is given by $\eta\textbf{\textit{j}}$ where $\eta$ is the 
resistivity and $\textbf{\textit{j}} = j_{z}\textbf{\textit{e}}_{z}$ is the electric current in the 2D 
case.

 \begin{figure}[t]
   \centering
    \includegraphics[width=\hsize]{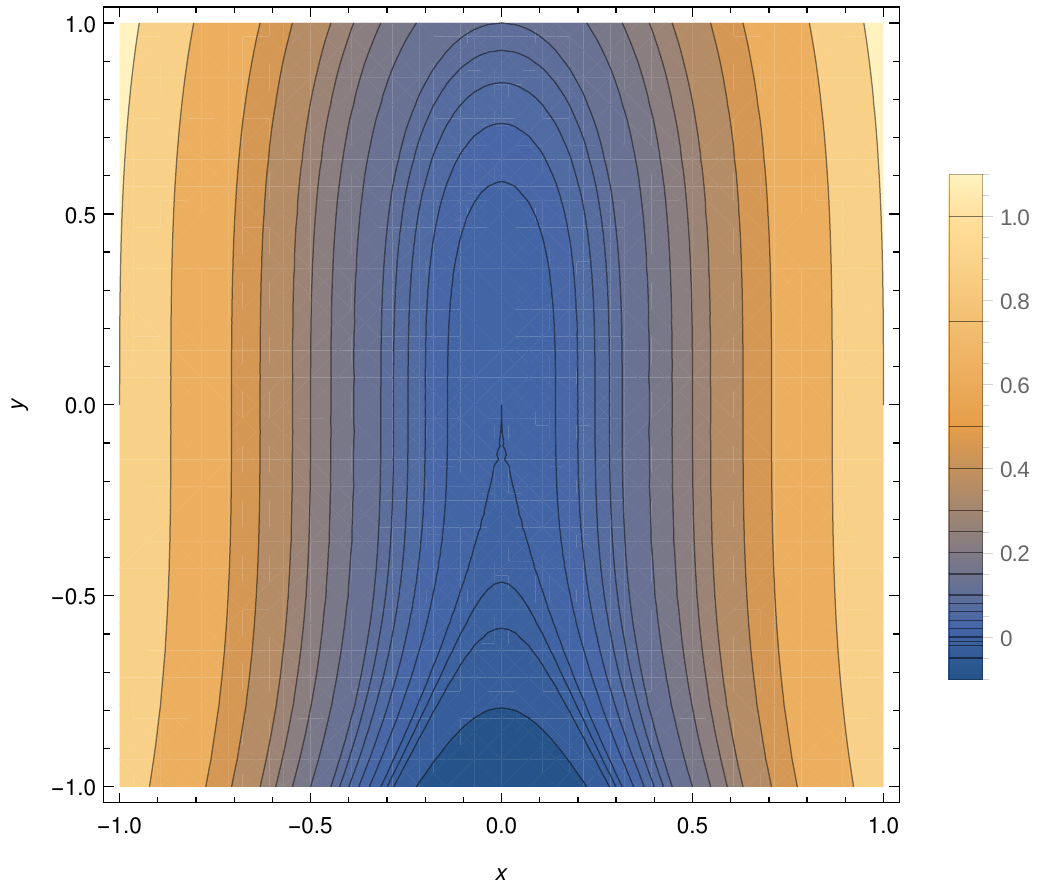}
      \caption{Field lines plotted as contour lines of the magnetic flux function $A$ defined by Equation~(\ref{Adef}) and for $a=-0.1$. 
      }
         \label{fig:cusp}
  \end{figure}
  
In the following we restrict the analysis to stationary flows. More specifically, we concentrate on the 
investigation of non-ideal Ohm's law, as it reflects magnetic topology and magnetic flux conservation
and their violation. Furthermore, we focus on cusp configurations which are structurally unstable but 
topological classifiable magnetic singularities. For structurally stable magnetic null points an 
indispensable criterion for magnetic reconnection is that the plasma flow can cross magnetic 
separatrices, where the plasma flow is usually connected to an X-type stagnation point flow. Cusp 
structures are thought to result from reconnection processes. However, as in the strict mathematical 
sense no magnetic separatrices exist at or near cusps, the question arises how the plasma streams in the 
vicinity of such structures.

\begin{figure*}[t]
\centering
\includegraphics[width=0.42\textwidth]{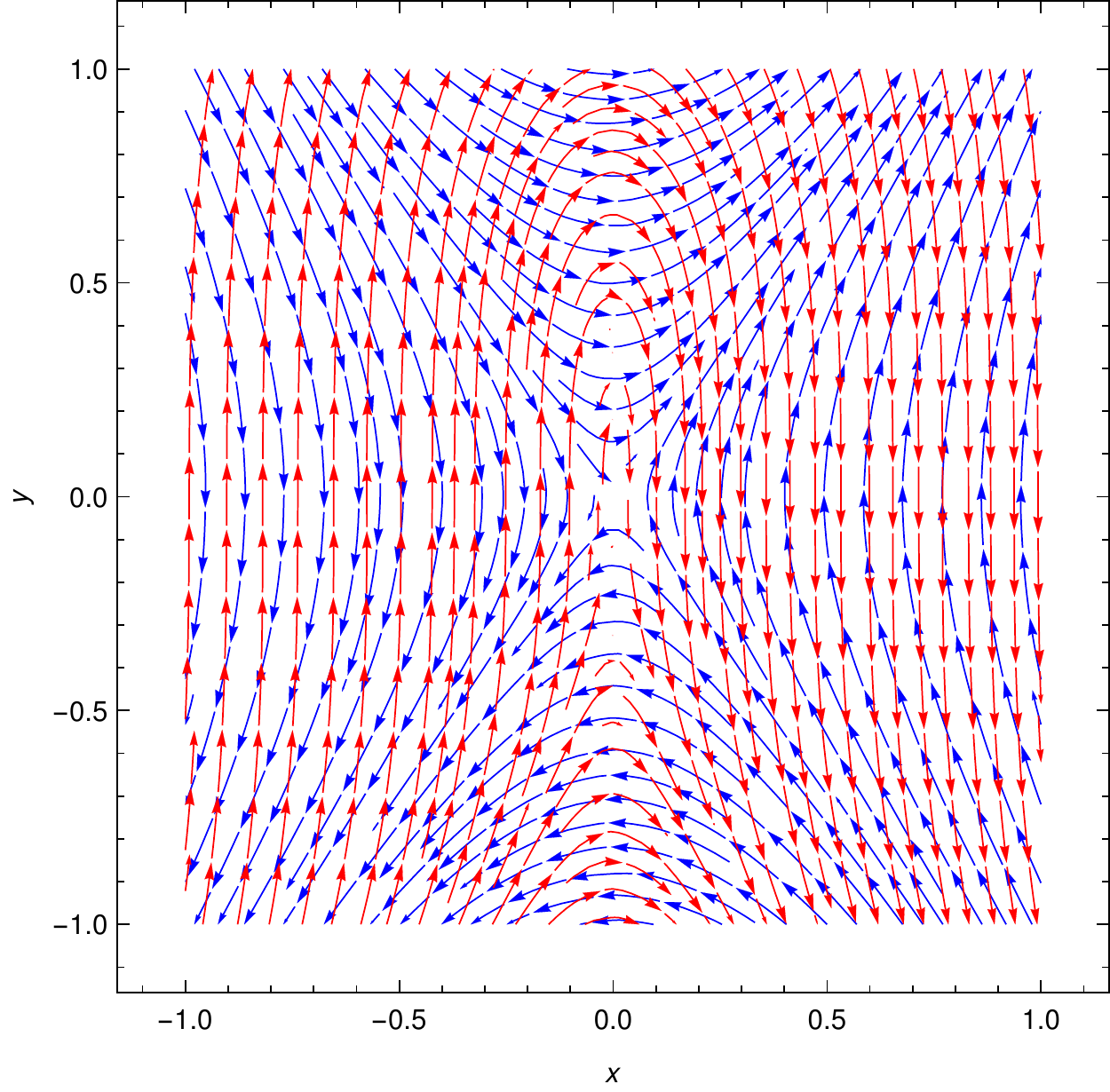}
\includegraphics[width=0.495\textwidth]{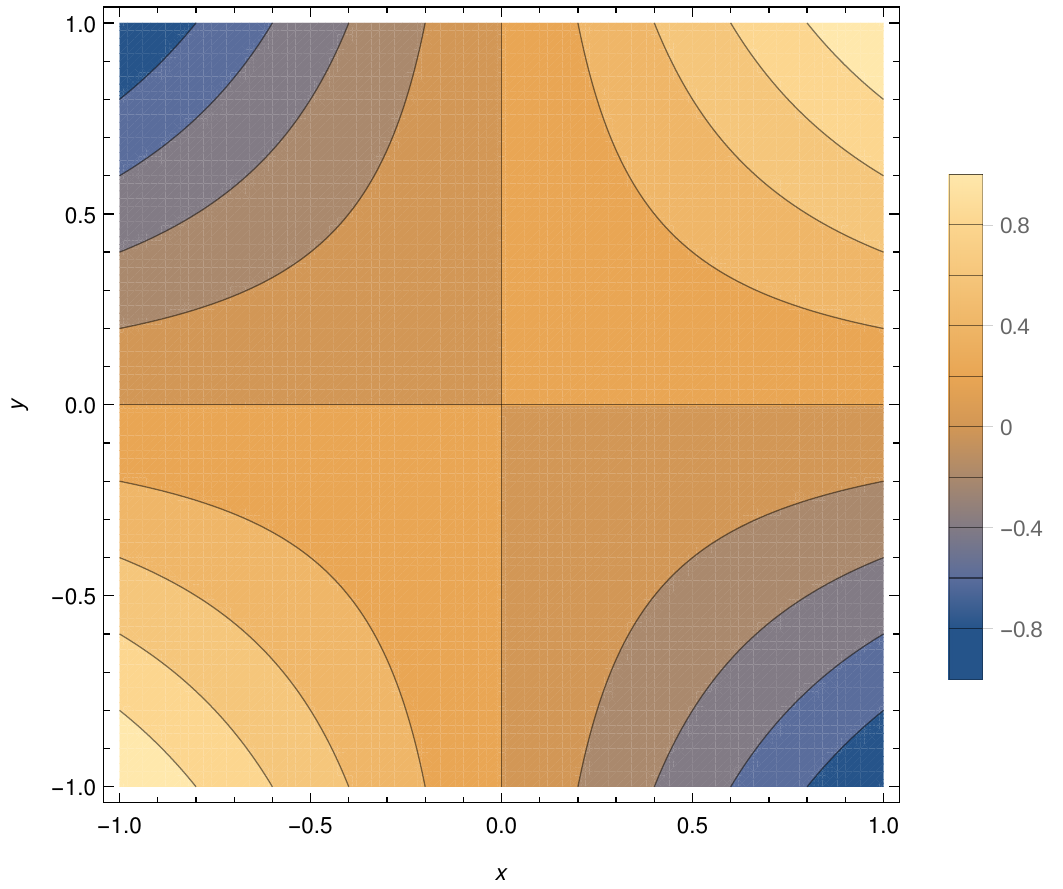}
\caption{{\it Left:} Magnetic cusp, represented by the magnetic field lines (red arrows), and streamlines (blue arrows). {\it Right:} Resistivity map. The calculations are for $E_z = 0$.}
\label{fig:Ez0}
\end{figure*}

To describe such flows, it is necessary to find solutions of the non-ideal Ohm's law, meaning that 
the plasma flow and the non-idealness are interconnected. Therefore, we need to find the connection 
between the topological classes of the vector fields, namely the plasma flow and the magnetic field,  
and the topological classes of the resistivity, as depicted in the following diagram.
\begin{eqnarray}
&&\boldsymbol\nabla\textbf{\textit{B}}\,\,\,\longleftrightarrow\,\,\, \boldsymbol\nabla\textbf{\textit{v}}\nonumber\\
&&\,\,\,\,\,\nwarrow\!\!\!\!\!\!\searrow\quad\quad\quad\nearrow\!\!\!\!\!\!\swarrow\nonumber\\
&&\,\,\,\boldsymbol\nabla\textbf{\textit{N}}
\stackrel{\textrm{2D}}{\widehat{=}} \boldsymbol\nabla \eta\, ,
\boldsymbol\nabla \boldsymbol\nabla \eta \nonumber
\end{eqnarray}
The quantitative indicators of the Poincar\'{e} class of the non-idealness are the eigenvalues of $\boldsymbol\nabla\textbf{\textit{N}}$. In 2D, we have to determine the topological indicators 
by analyzing the first and second derivatives of the resistivity. In the most general case,
the non-idealness contains also the electric current, $j_z$, which itself is a function of at least one spatial coordinate, here $y$, in the case of a cusp. This means that the polynomial functions 
describing $N_z$ are reducible by the polynomial factor given by $j_z$.

From the perspective of a micro-physical 2D approach, we have to take only non-ideal terms into 
account which point into the 
$z$-direction. To find a non-ideal term which can be described by a resistive interaction,
it is convenient to define a resistivity $\eta$, which is based on a collision frequency $\nu$, respectively a connected time-scale, being coupled to the 2D total electric field $E$ (respectively $N_{z}$).
This drift-velocity ansatz uses a frictional force $\nu v_{D}$ with $v_{D}$ as the drift velocity between electrons and ions. Assuming stationarity, $d v_{D}/dt = 0$, and using the definition of the electric current density,
$j_{z}= q n v_{D}$ with the particle density $n$ in the 
quasi-neutral approximation and the particle charge $q$, it follows
\begin{equation}
\frac{d v_{D}}{dt} =\frac{q}{m}\, E - \nu v_{D} = 0 \quad\land\quad E = \eta j_{z}=\eta q n v_{D}\, ,
\end{equation}
where  $m$ is the mass of the particle. 
Consequently, we find for $\eta$
\begin{equation}
\quad \eta = \frac{m \nu}{n q^2}\, .
\end{equation}
For every 2D non-ideal term such a resistivity can obviously be calculated. 

However, our focus is not 
on the specific micro- and macro-physical mechanisms causing resistivity, but on the question which 
topological requirements the function $\eta(x,y)$ must fulfill in the frame of MHD, independent of the 
explicit macro-physical parameters. For this reason, we want to investigate how flows and electric 
fields and assigned resistivities or non-ideal terms can 
be topologically characterized around magnetic cusp structures, using the classification introduced by 
\citet{hp1881jm}, \citet{arnold}, and \citet{Lyapunov}.

In the next section we show how fields and functions can 
be classified and algebraically represented, using the example for the magnetic field with cusp 
structure.

\subsection{Topological classification of cusp-type magnetic null points by $A_k$-type singularities}
 
We investigate the structure of degenerated magnetic flux functions $A$ for which null points of 
cusp-type of 
lowest order exist. Such null points are formally structural unstable, i.e. degenerated in the sense of 
the so-called \lq $A_{k}$-singularities\rq~defined by \citet{arnold}. In mathematics, especially in the 
theory of singularities, $A_{k}$ with $k\ge 0$ describes the level of degeneration of a function. 
Non-degenerated functions, for example the Morse functions which are locally, in the vicinity of a null 
point, represented by polynomials of second degree, have the degeneration level $k = 0$. For $k = 2$, 
the resulting degeneracy contains locally a cusp structure.

Following this approach, all regular 2D cusp-type magnetic fields of lowest order ($k=2$) can be 
canonically represented by the magnetic flux function
\begin{equation}
A=x^2-a y^{3}\, , \label{Adef}
\end{equation}
whereby we introduced a scale factor given by the parameter $a$, which regulates the width and 
elongation of the cusp. This flux function displays for $A=0$ a field line which is a Neil's parabola, 
containing a cusp of lowest order, a so-called spinode, which is located at the origin (see 
Figure~\ref{fig:cusp}).

The magnetic field can be computed 
according to $\textbf{\textit{B}} = \boldsymbol\nabla A \times \textbf{\textit{e}}_{z}$, and the electric current is given by $-\Delta A = j_{z} = -2+6ay$.

\begin{figure*}[t]
\centering
\includegraphics[width=0.42\textwidth]{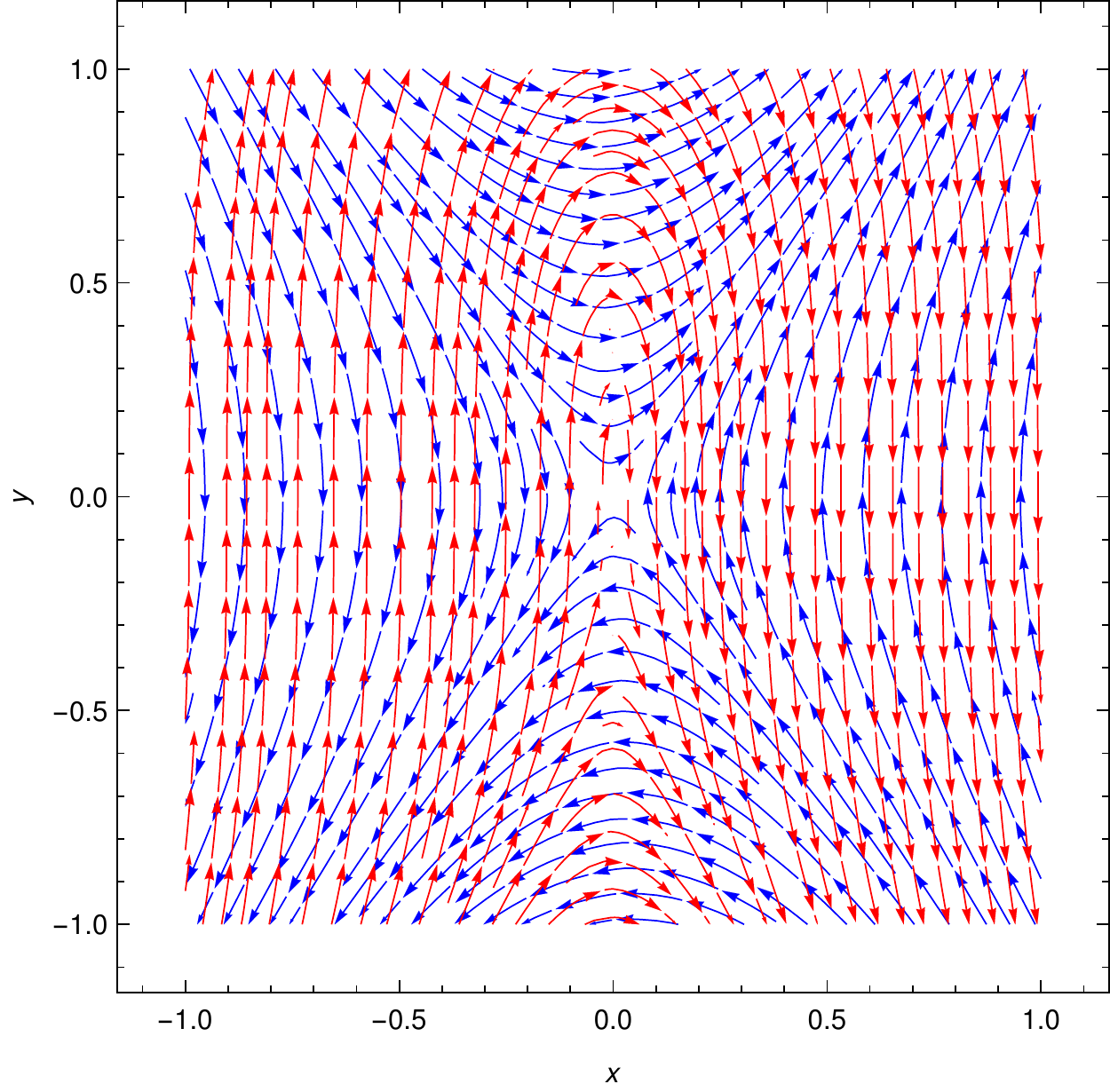}
\includegraphics[width=0.5\textwidth]{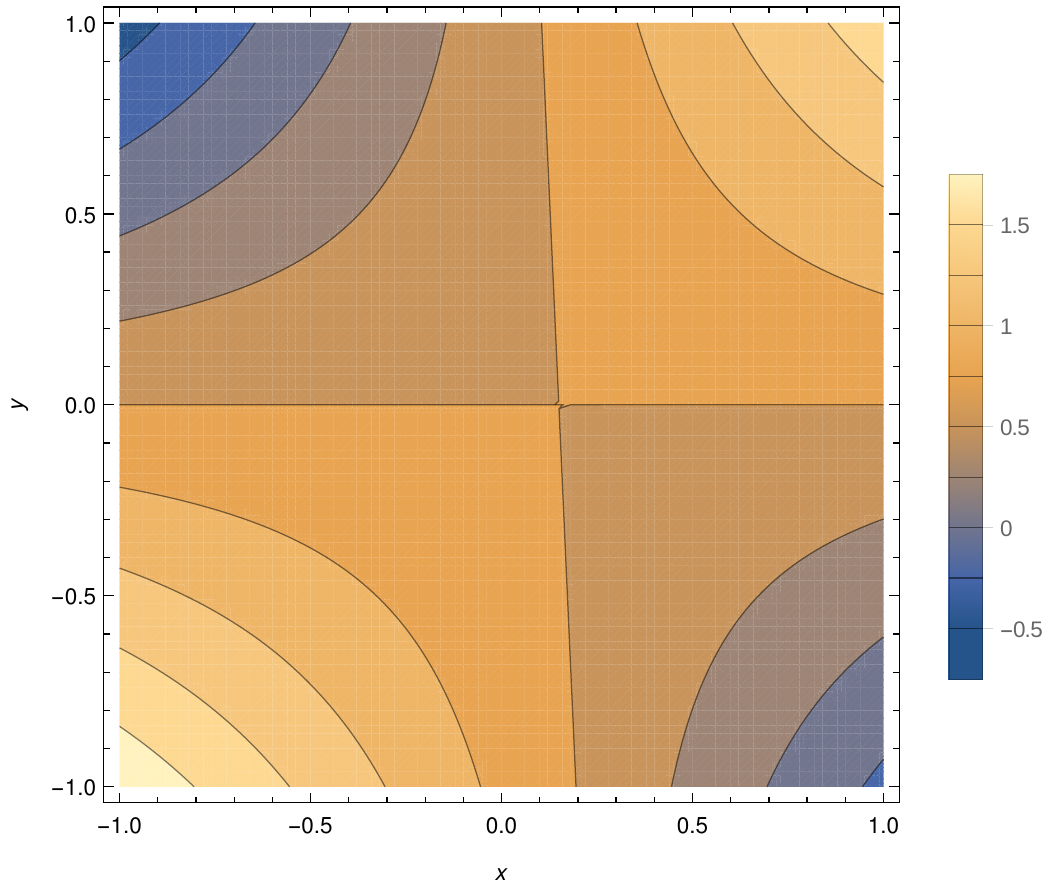}

\includegraphics[width=0.42\textwidth]{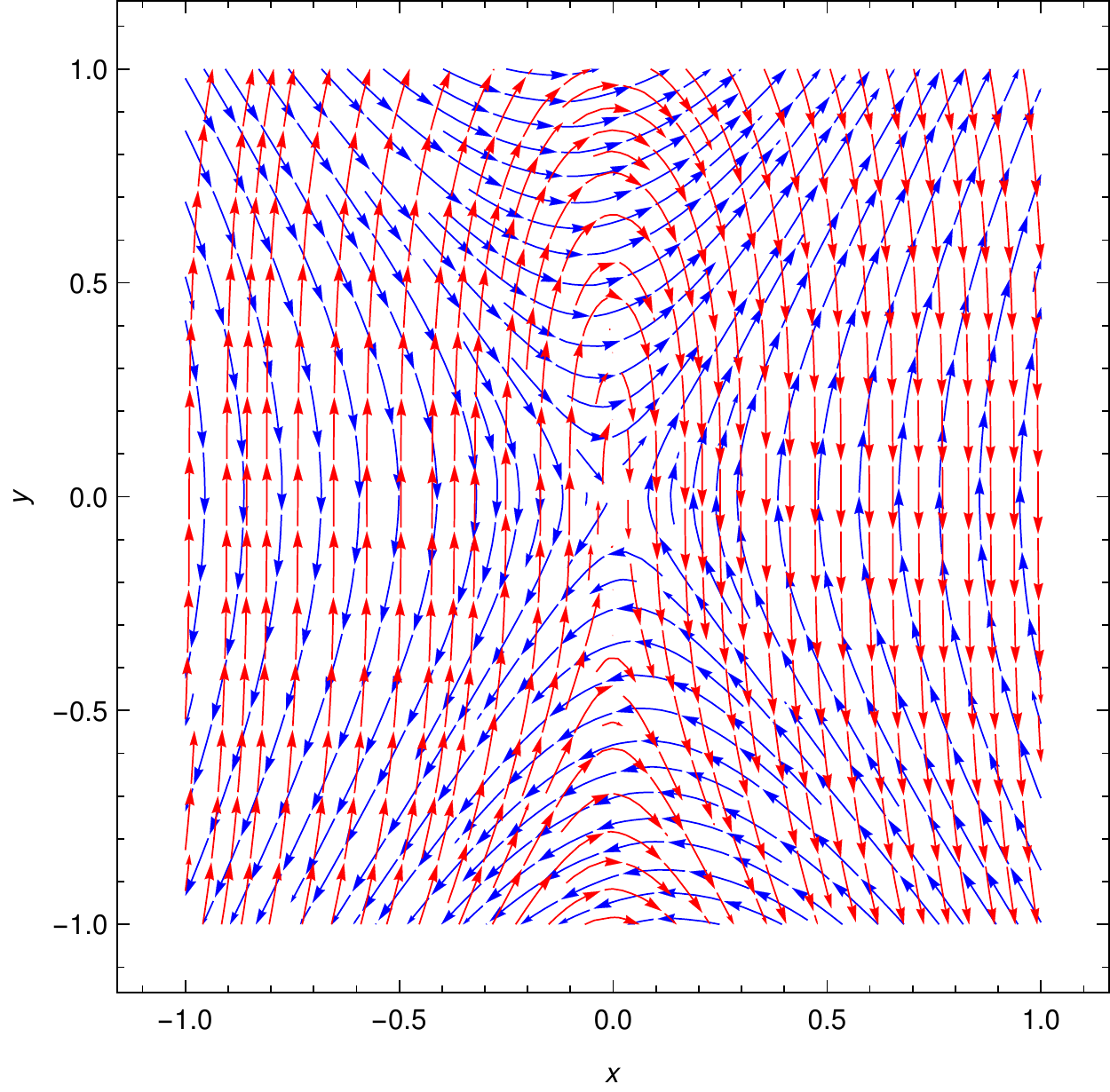}
\includegraphics[width=0.49\textwidth]{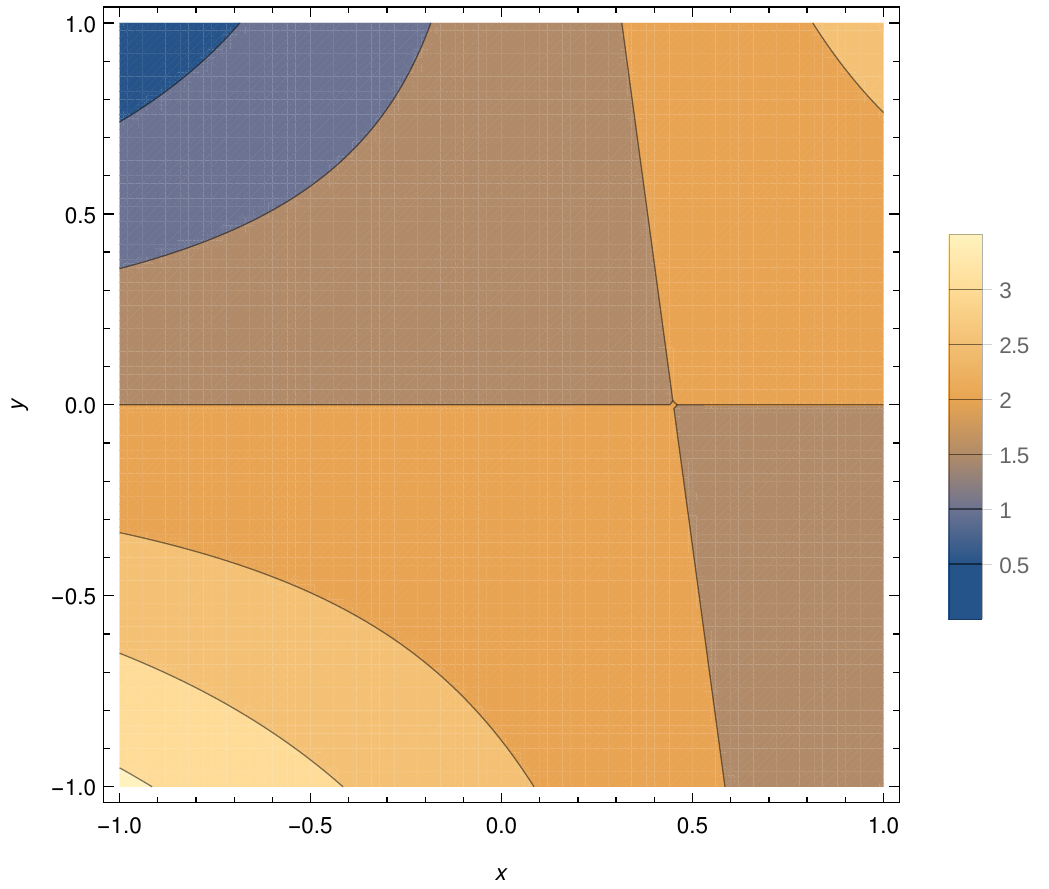}

\includegraphics[width=0.42\textwidth]{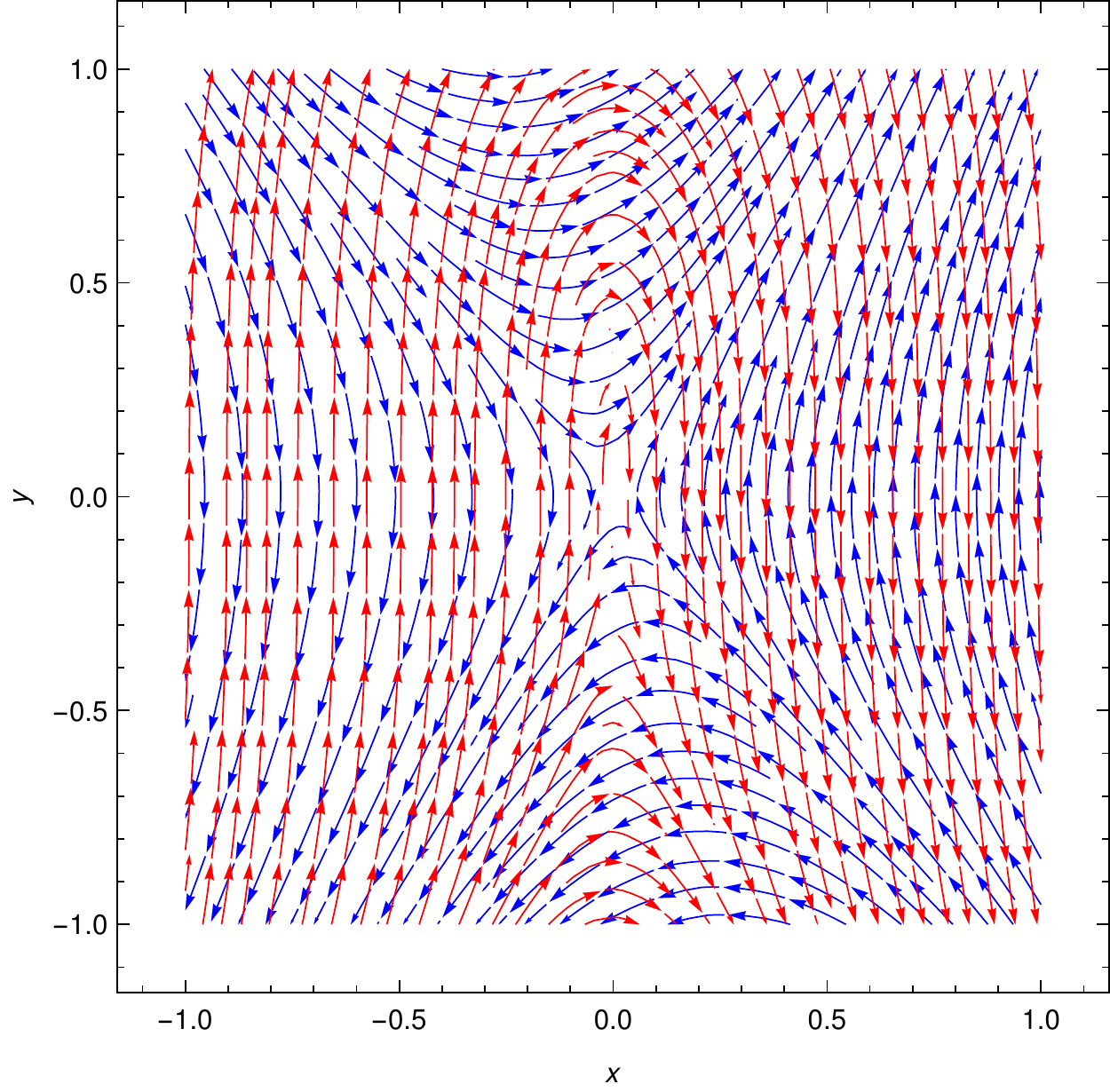}
\includegraphics[width=0.48\textwidth]{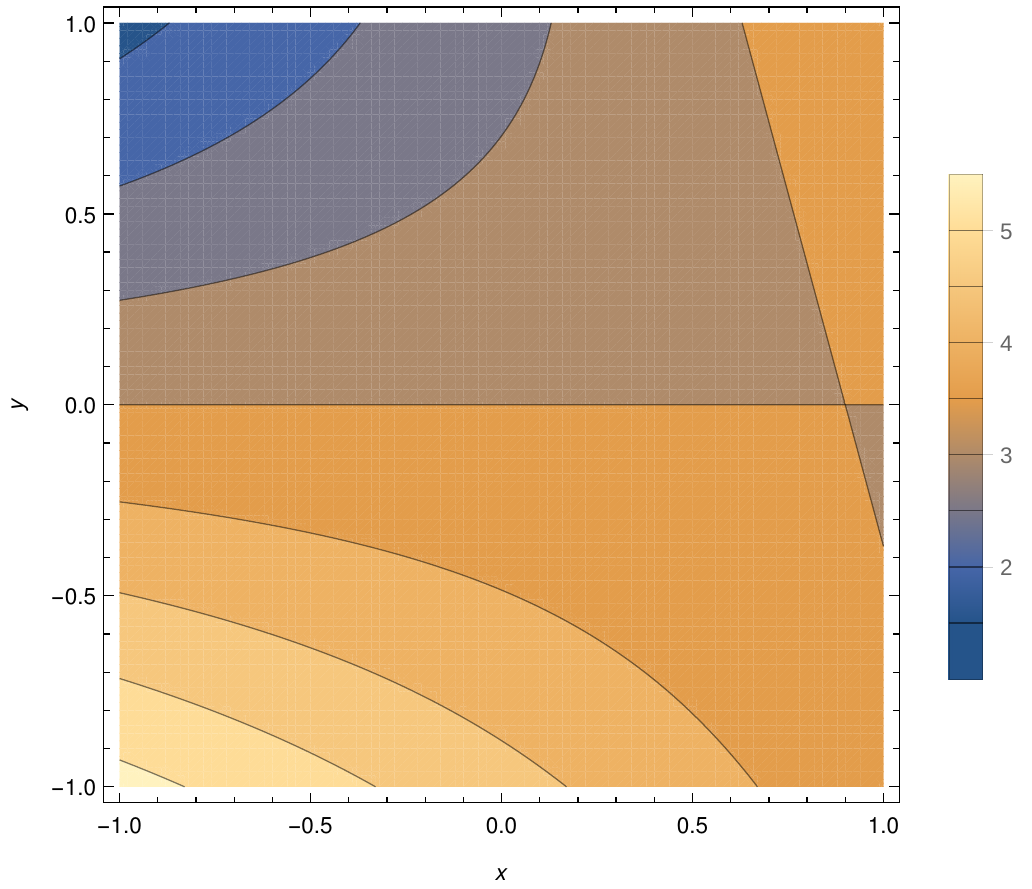}
\caption{{\it Left:} Depiction of the magnetic field lines (red arrows) and the streamlines (blue arrows). {\it Right:} Resistivity maps. The calculations are for $E_z = -1$ (top), $E_z = -3$ (middle),
and $E_z = -6$ (bottom) and fixed value of $V_{12} = 1$.}
\label{fig:Ezneq0}
\end{figure*}

\section{Results}

We investigate a stationary stagnation-point flow in the vicinity of a magnetic cusp.
This is an analogy to reconnection flows in the vicinity of a non-degenerated regular magnetic null 
point like an X-point.
The analysis is restricted to
the non-ideal Ohm's law, as the corresponding relations between the eigenvalues of the gradients of
magnetic field, velocity field and non-idealness are determined by the flux transport, given
by a slightly violated ideal Ohm's law.
To allow for a deviation or relaxation of the classical reconnection affine like stagnation-point flow,
we assume that the stagnation point of the flow can be displaced with respect to the
origin, i.e., the location of the cusp. The ansatz for the velocity field is therefore
\begin{eqnarray}
v_{x} &&=V_{10}+ V_{11}x+V_{12}y\\
v_{y} &&=V_{20}+V_{21}x+V_{22}y\, ,
\end{eqnarray}
where  $V_{11}$ to $V_{22}$ are the  components of the velocity gradient, and $V_{10}$ and $V_{20}$ 
are the velocity displacements in $x$ and $y$ direction, respectively. The ansatz for the resistivity 
is given by the quadric 
\begin{eqnarray}
\eta=\eta_{00}+\eta_{10}x+\eta_{01}y+\eta_{20}x^2+\eta_{02}y^2+\eta_{11}xy\, , \label{etadef}
\end{eqnarray}
where the $\eta_{ik}$ describe the components of the Hessian ($\eta_{20}$, $\eta_{02}$, and $\eta_{11}$) 
and the Jacobian matrix ($\eta_{10}$ and $\eta_{01}$) of the resistivity, and the value of the 
resistivity at the null point ($\eta_{00}$). All coefficients of $v_{ij}$ and $\eta_{ij}$ are 
basically unknown and correlated by the Poincar\'{e} class of the magnetic field and the MHD equations.
As we are searching for a general solution for all possible $v_{ij}$ and $\eta_{ij}$, it is not useful
to prescribe and fix $\eta$ as is typically done in analytical or numerical computations.

Inserting Equations~(\ref{Adef})-(\ref{etadef}) into the non-ideal Ohm's law Equation~(\ref{ohm1}), sorting by the orders of the monomials in $x$ and $y$, and comparing the coefficients delivers
\begin{eqnarray}
E_z &=& -2\eta_{00}\label{or0}\\
 -2 V_{10} &=&-2\eta_{10}\label{or1} \\ 
0 &=& 6a\eta_{00}-2\eta_{01}\label{or2}\\
-2V_{12}&=&  6a\eta_{10}-2\eta_{11}\label{or3}\\
  -2 V_{11} &=& -2\eta_{20}\label{or4}\\
3a V_{20} &=& 6a\eta_{01}-2\eta_{02}\label{or5}\\
 0 &=& 6a\eta_{20}\label{or6}\\
 3aV_{21}&=& 6a\eta_{11}\label{or7}\\
3a V_{22} &=&  6a\eta_{02}\label{or8}\, .
\end{eqnarray}
We fix the constant component $E_z$ and solve for the other coefficients. We obtain the following
relations
\begin{eqnarray}
\eta_{00}& =& -\frac{1}{2}\, E_{z}\\
\eta_{20}&=&0 \\
\eta_{01}& = &-\frac{3}{2}\,a E_{z}\\
V_{11} & = &0\\
V_{10} & = & \eta_{10}\\
 -  V_{12} & = &  3a\eta_{10}-\eta_{11}\\
 3a V_{20} &=& -9 a^2 E_{z}-2\eta_{02}\\
    V_{21} &=&  2\eta_{11}\\
    V_{22} &=&  2\eta_{02}\, .
\end{eqnarray}
%
The reconnecting electric field $E_{z}$  should be coupled mainly to the resistive term, including 
resistivity and current. To minimize the flux transport by a constant velocity field ($V_{10}$, 
$V_{20}$), we set the displacement terms to zero, $V_{10}=V_{20}=0$. This choice also guarantees that  
the singular magnetic null point is connected with the singular point of the flow.
Then the equations simplify to 
\begin{eqnarray}
\eta_{00}& =& -\frac{1}{2}\, E_{z} \label{set_eta1}\\
 \eta_{10}& = &0\\
 \eta_{20}& =& 0\\
\eta_{01} &=& -\frac{3}{2}\,a E_{z}\\
\eta_{02}&=&-\frac{9}{2} a^2 E_{z}\\
 V_{11}&=&0\\
 V_{12}&=&\eta_{11}\\
 V_{21}&=&2V_{12}=2\eta_{11}\\
 V_{22}&=&-9a^2 E_{z}\, . \label{set_eta10}
\end{eqnarray}

Next, we calculate the determinant of the Jacobian matrix of $\textbf{\textit{v}}$ 
\begin{equation}
\left|
\begin{array}{cc}
0-\lambda_{v} & V_{12}\\
2V_{12} & V_{22}-\lambda_{v}\\
\end{array}
\right|=\lambda_{v}^2-V_{22}\lambda_{v}-2V_{12}^2=0
\label{char2}
\end{equation}
%
The eigenvalues are then given by the solutions of the characteristic 
equation Equation~(\ref{char2})
\begin{eqnarray}
\lambda_{v} &=&\frac{V_{22}}{2}\,\pm\,\sqrt{\frac{1}{4} V_{22}^2+2V_{12}^2}\\
&=& -\frac{9a^2 E_{z}}{2}\,\pm\,\sqrt{\frac{81}{4} a^4 E_{z}^2+2\eta_{11}^2}\, .
\end{eqnarray}
These eigenvalues provide us with the Poincar\'{e} class of the velocity field, which must be 
hyperbolic in the studied case, meaning that the flows are in general sheared hyperbolic X-type 
structures with a separatrix consisting of two branches given by
\begin{equation}
y = \frac{\lambda_v}{V_{12}}\, x = \frac{-\frac{9a^2 E_{z}}{2}\,\pm\,\sqrt{\frac{81}{4} a^4 E_{z}^2+2V_{12}^2}}{V_{12}}\,x\, .
\end{equation}
These define two straight lines (for $V_{12} \neq 0$) intersecting at the origin. If the flow is irrotational, i.e. 
$V_{12} = \eta_{11} = 0$, one eigenvalue is zero and the other one is $\lambda_v = -9 a^2 E_z$.
This case is degenerated because the flow turns into a 
one-dimensional shear-flow, the separatrix vanishes, and instead of a null point a null line of the 
flow occurs. The flow then points into $y$-direction and depends only linearly on $y$. 

\begin{figure*}[t]
\centering
\includegraphics[width=0.42\textwidth]{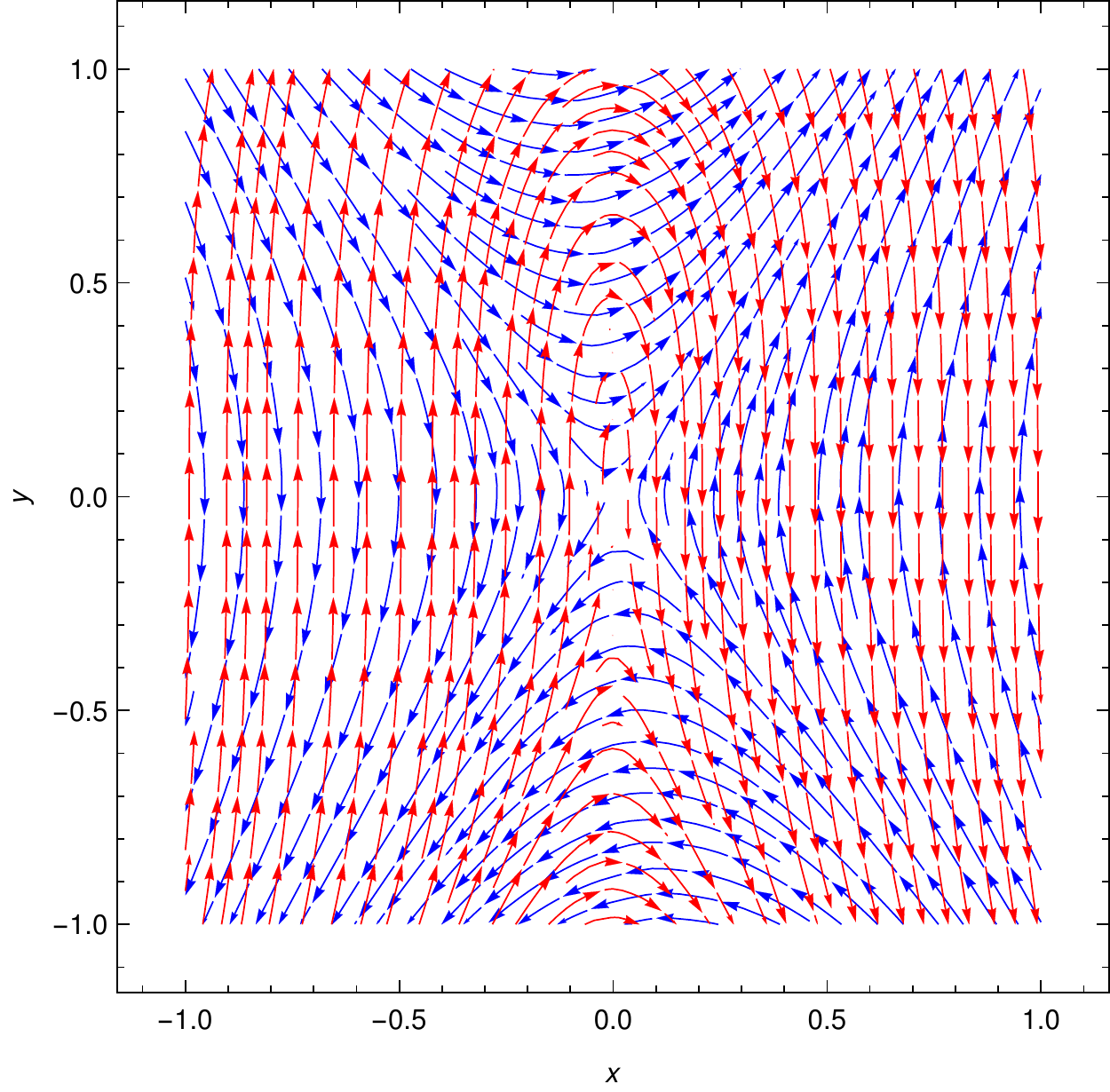}
\includegraphics[width=0.48\textwidth]{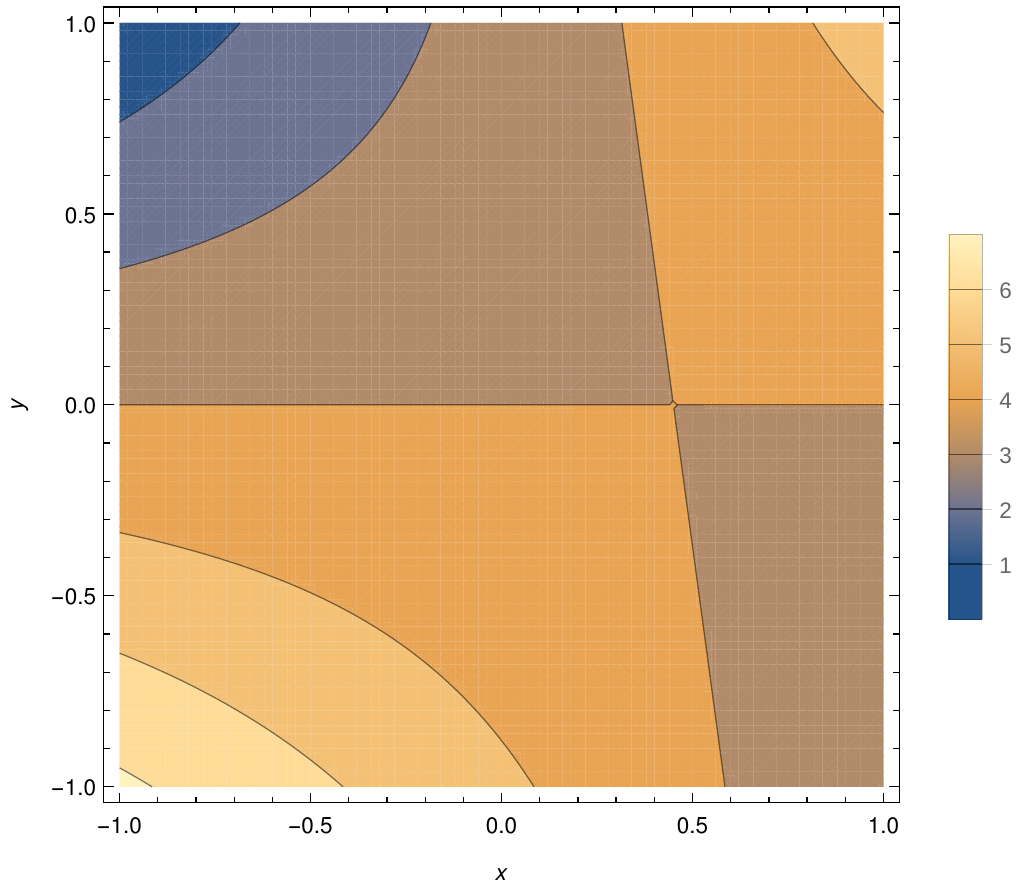}

\includegraphics[width=0.42\textwidth]{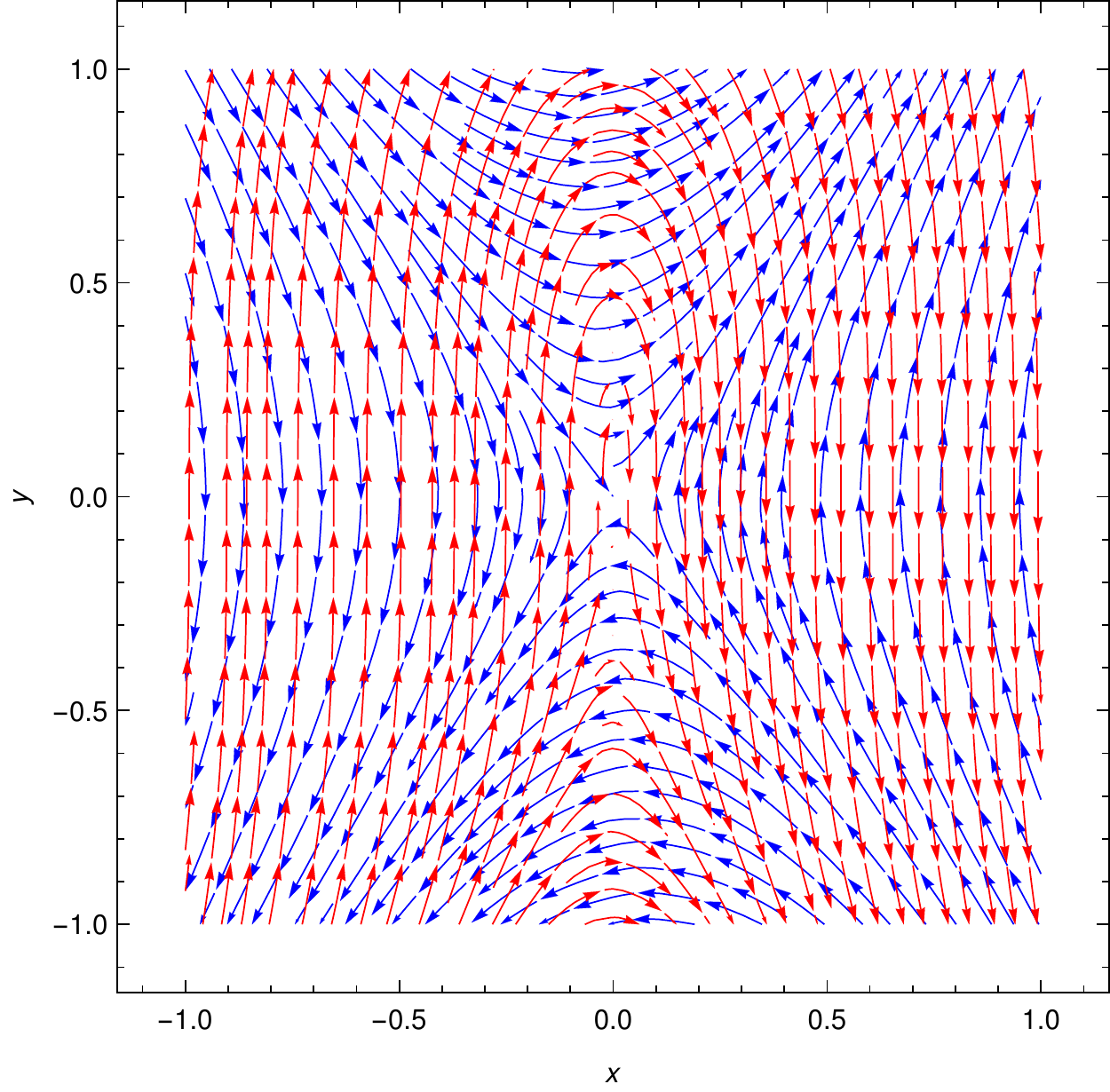}
\includegraphics[width=0.48\textwidth]{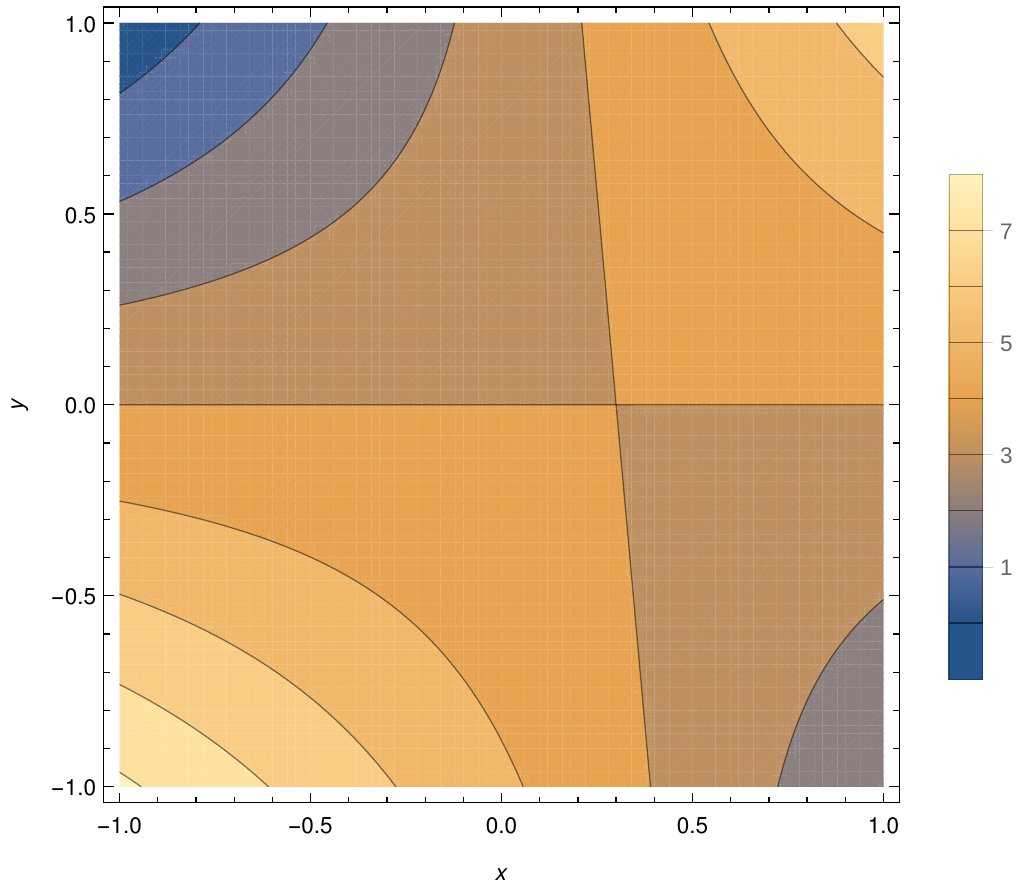}

\includegraphics[width=0.42\textwidth]{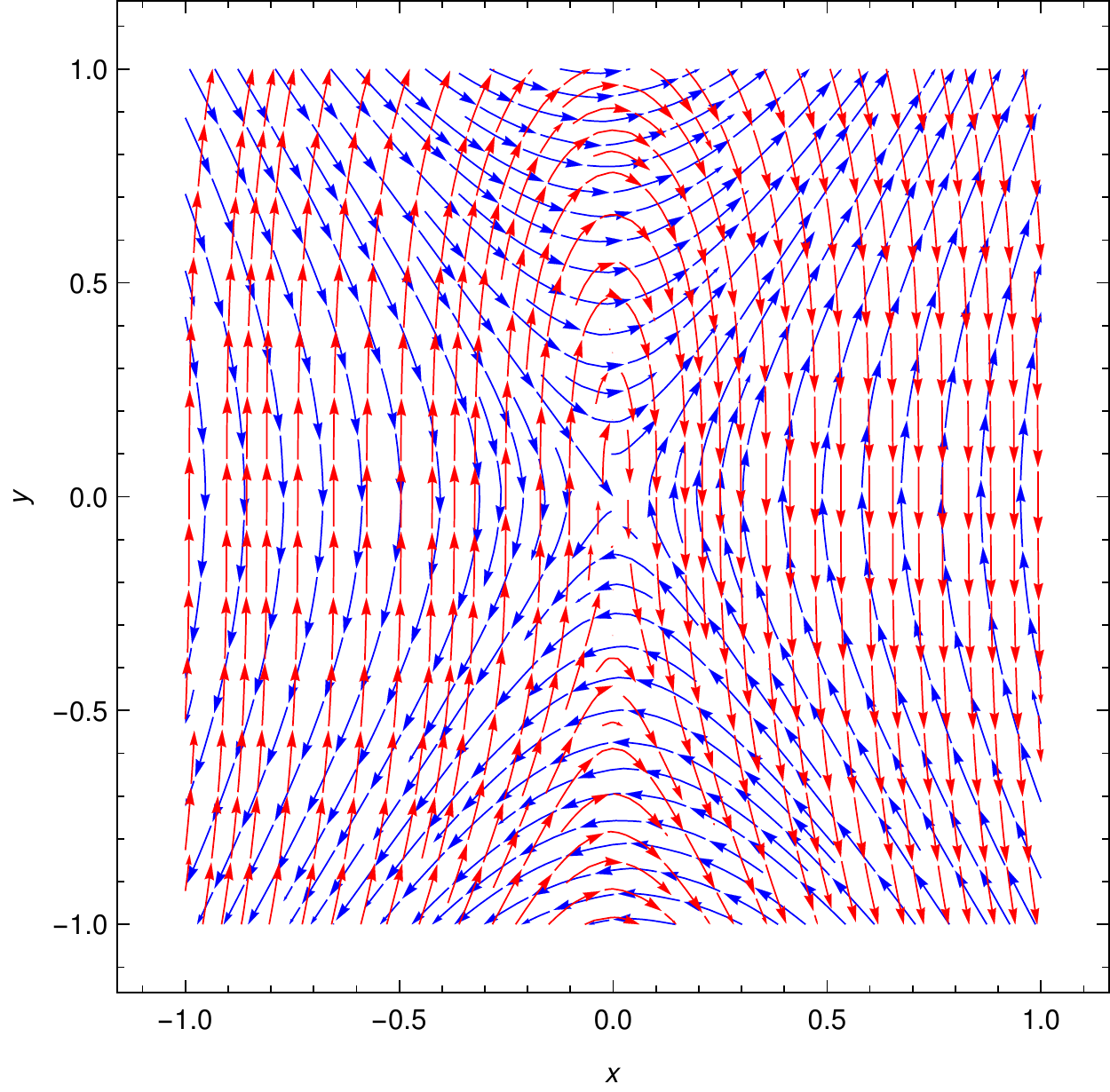}
\includegraphics[width=0.49\textwidth]{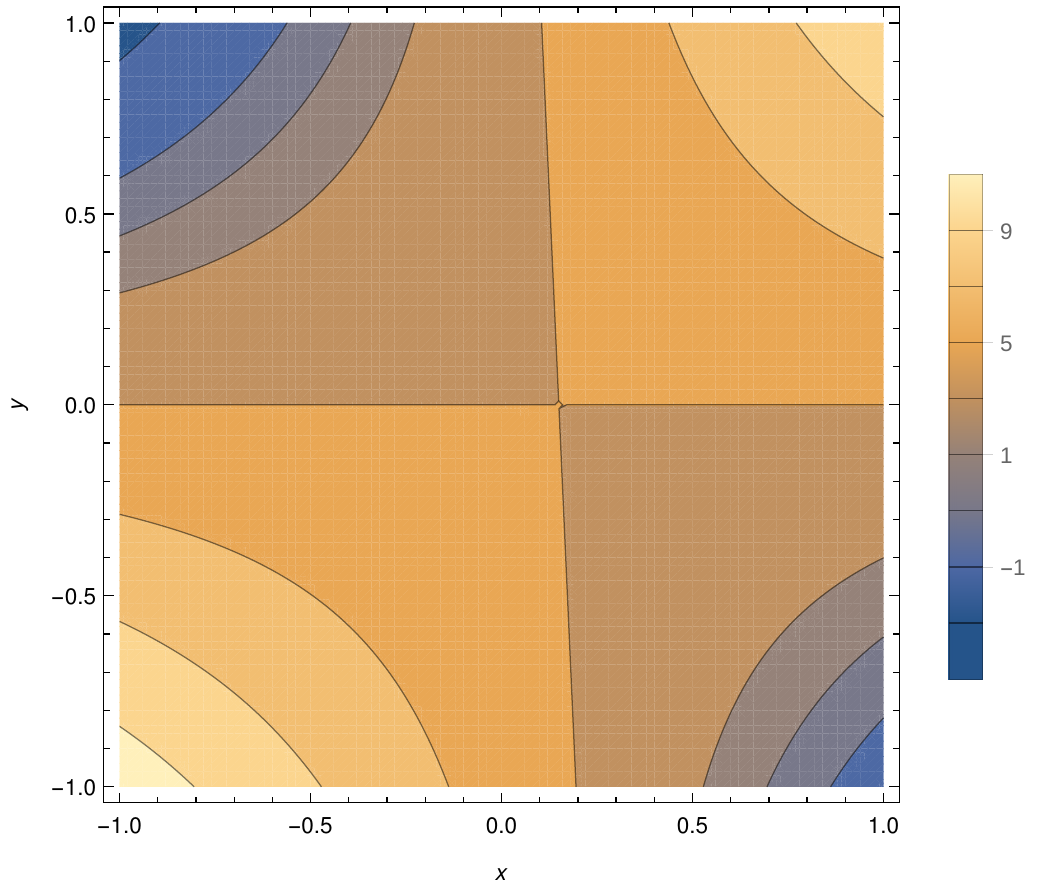}
\caption{{\it Left:} Depiction of the magnetic field lines (red arrows) and the streamlines (blue 
arrows). {\it Right:} Resistivity maps. The calculations are for a fixed value of $E_z = -6$ and for 
values of $V_{12} = 2$ (top), $V_{12} = 3$ (middle), and $V_{12} = 6$ (bottom).}
\label{fig:V12}
\end{figure*}

Now we turn to the resistivity. From inserting the set of Equations (\ref{set_eta1}) - 
(\ref{set_eta10}) into Equation (\ref{etadef}), we find that its quadric takes the following form
\begin{eqnarray}
f(x,y,\eta)&=&-\eta-\frac{1}{2}\, E_{z}-\frac{3aE_{z}}{2}\, y -\frac{9a^2E_{z}}{2}\, y^2 + 
V_{12}xy \nonumber\\
& = & 0\, .
\label{eta} 
\end{eqnarray}
This equation defines the spatially varying resistivity $\eta$.
Any quadric has three invariants in the form of determinants
\citep[see, e.g.,][]{bartsch} which, in our case, are given by the following
determinants for $\Delta$, $\delta$, and $t$:

\begin{equation}
\Delta =
\left|
\begin{array}{cccc}
                 0 & \frac{1}{2} V_{12}    & 0            & 0\\
\frac{1}{2} V_{12} & -\frac{9a^2E_{z}}{2}  & 0            & -\frac{3aE_{z}}{2}\\
                 0 &                  0    & 0            & -\frac{1}{2} \\
                 0 & -\frac{3aE_{z}}{2}    & -\frac{1}{2} & -\frac{1}{2} E_{z} \\
\end{array} 
\right|
= \frac{V_{12}^{2}}{4} > 0\, ,
\label{grdelta}
\end{equation}

\begin{equation}
\delta=
\left|
\begin{array}{ccc}
                 0 & \frac{1}{2} V_{12}    & 0      \\
\frac{1}{2} V_{12} & -\frac{9a^2E_{z}}{2}  & 0      \\
                 0 &                  0    & 0      \\
\end{array}
\right|
=0\, ,
\end{equation}

\begin{equation}
  t = 
\left|
\begin{array}{cc}
                 0 & \frac{1}{2} V_{12}  \\
\frac{1}{2} V_{12} & -\frac{9a^2E_{z}}{2} \\
\end{array}
\right|
+
\left|
\begin{array}{cc}
-\frac{9aE_{z}}{2}  &  0  \\
                0   &  0  \\
  \end{array}
\right| \nonumber\\
+
\left|
\begin{array}{ccc}
  0 &  0 \\
  0 &  0 \\
\end{array}
\right|
= -\frac{V_{12}^{2}}{4}\, .
\end{equation}
The results
\begin{equation}
\Delta>0 \quad\textrm{and}\quad t<0
\label{grdelta1}\end{equation}
together with 
\begin{equation}
\delta=0 
\label{kldelta1}
\end{equation}
lead to the implication that the resistivity surface above the ($x,y$)-plane is a hyperbolic
paraboloid. This means that the resistivity has a saddle-point structure in the vicinity of the 
magnetic singularity. Moreover, the absence of a term with $x^2$ in the description of the resistivity 
(Equation~(\ref{eta})) implies that the saddle point is located on the $x$-axis.

To illustrate this behavior, we compute the magnetic and flow fields as well as the resistivity map. 
Free parameters of our model are $a$, $V_{12}$, and $E_{z}$. The parameter $a$ controls
the width and elongation of the cusp, and we fix it at $a=-0.1$. For the velocity component $V_{12}$ we 
use a value of $V_{12} = 1$ in normalized units. To guarantee that the resistivity is positive 
in the vicinity of the cusp point, we need to use a suitable value for $E_z$. This value needs to be 
$<0$. In the limiting case of $E_z=0$ the resistivity at the cusp vanishes. This is demonstrated in the
right panel of Figure~\ref{fig:Ez0}, where the resistivity is zero along the $x$ and $y$-axes, positive 
in the first and third quadrants, and negative in the other two. The magnetic cusp structure is shown
via the magnetic field lines (red arrows) in the left panel of Figure~\ref{fig:Ez0}. In this plot, we 
overlaid the streamlines (blue arrows), which are symmetric with respect to the X-point at the origin.

For this limiting case of $E_z = 0$, representing an incompressible flow, the convective flux 
transport by $\textbf{\textit{v}}\times\textbf{\textit{B}}$  can only be compensated by the resistive 
term. As due to the hyperbolic structure of the flow magnetic flux is transported in and out of the cusp 
region, only in 2 quadrants magnetic flux can be effectively annihilated, whereas in the other two 
magnetic flux needs to be built up. It is known that the electric field component in the rest frame,
$E_z$, reflects the strength of the total flux transport and should hence not vanish. So the larger 
$|E_z|$, the more magnetic flux can be annihilated. Otherwise the transport of magnetic flux reflects 
some kind of dynamo process, which requires a negative resistivity. This transport of magnetic 
flux implies a conversion from kinetic into magnetic energy, which means that 
energy needs to be supplied to the system instead of dissipation taking place.

For the general, compressible case with non-vanishing $E_z$, the two eigenvalues $\lambda_v$ are 
asymmetric and, therefore, reflect an X-point configuration with separatrix branches which are not 
symmetric anymore with respect to the $y$-axis. For fixed values of $a$ and $V_{12}$ we find that the 
more negative the value of $E_z$, the steeper the separatrix branch with the negative slope and the more 
horizontal the separatrix branch with the positive slope. This can be seen in the left panels of 
Figure~\ref{fig:Ezneq0} where we show the flow patterns for $E_z = -1$ (top), $E_z = -3$ (middle), and 
$E_z = -6$ (bottom). In the limiting case $|E_z| \rightarrow \infty$, the angle between 
both branches converges to 90\degr and the separatrices coincide with the $x$ and $y$-axes. 

The corresponding resistivity maps to the three computed models are displayed in the right panels of 
Figure~\ref{fig:Ezneq0}. We note that the more negative $E_z$, the absolute values of the resistivity 
contours increase. Moreover, the saddle point of the resistivity is displaced in positive $x$-direction. 
Its position with respect to the origin is given by
\begin{equation}
\delta x = -\frac{1}{2}\,\frac{3a|E_z|}{V_{12}}\, .
\end{equation}
The two resistivity separatrices result from
\begin{eqnarray}
y &= &0\, , \\
y &= &\frac{V_{12}}{9 a^2 E_z}\, x - \frac{1}{6a}\, .
\end{eqnarray}
Obviously, in the case of a fixed value for $V_{12}$, the slope of the second separatrix decreases with 
increasing value of $|E_z|$, as can be seen in right panels of Figure~\ref{fig:Ezneq0}.

The displacement of the saddle point is caused by the choice of the geometrical configuration for the 
magnetic and the flow field, in combination with the asymmetry of the electric current. This 
displacement can be regulated by the choice of the free parameters $a$, $E_z$, and $V_{12}$. To minimize 
the offset and hence to eliminate the asymmetry of the resistivity function, and also to reduce the 
compressibility of the flow, one might chose a value for $a$ that approaches zero. This, in turn, causes 
a decrease of the width and a steepening of the cusp structure. Alternatively, if the shape of the 
magnetic cusp should remain unchanged, we may fix $E_z$ and 
increase the value of $V_{12}$. This is shown in Figure~\ref{fig:V12}, where we use $E_z = -6$ and 
$V_{12}=2$ (top), $V_{12}=3$ (middle), and $V_{12}=6$ (bottom). The value of the resistivity at the 
saddle point does not depend on the choice of $V_{12}$, but the slope and offset of the second 
separatrix do. With increasing value of $V_{12}$, the slope increases and the offset decreases. In 
addition, the saddle structure of the resistivity steepens and its values considerably increase compared 
to the case with fixed $V_{12}$ and varying $E_z$. Moreover, an increase in $V_{12}$ counteracts 
the asymmetrization of the flow, as can be seen in the left panels of Figure~\ref{fig:V12}.

\begin{figure}[t]
\centering
\includegraphics[width=0.5\textwidth]{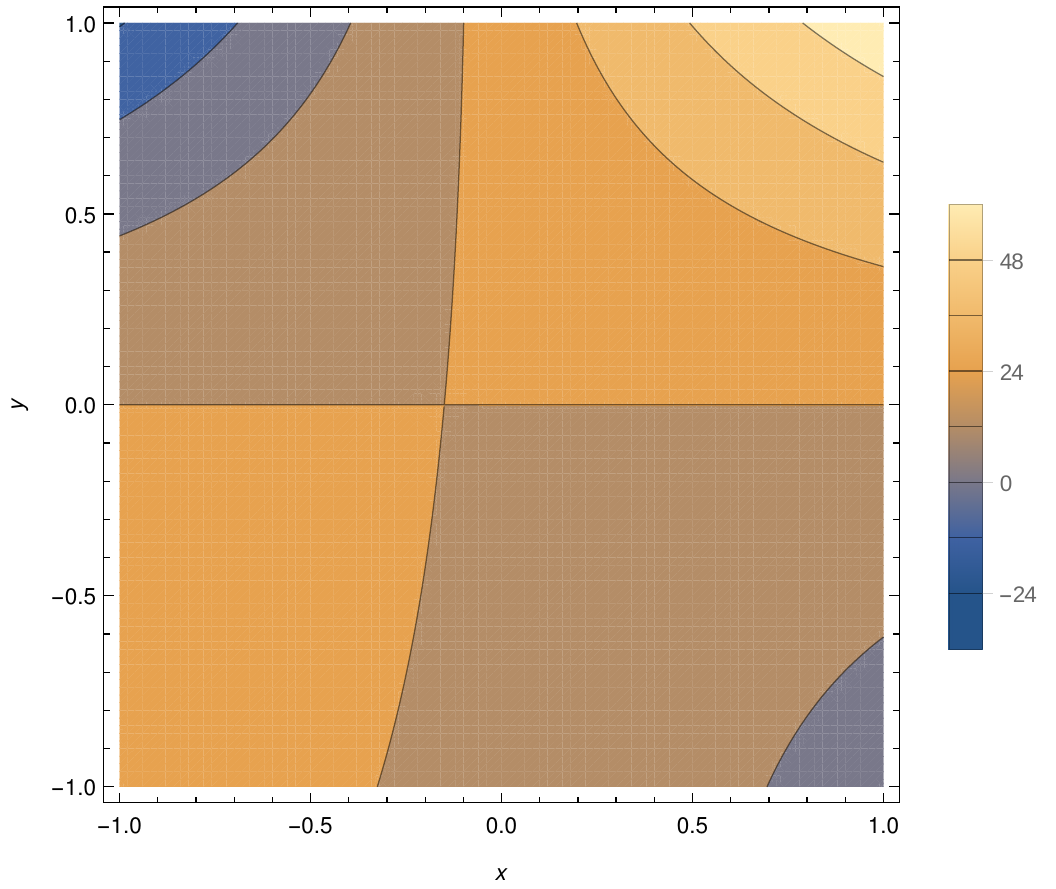}
\caption{Isocontours of the energy dissipation for $E_z = -6$ and $V_{12} = 6$.}
\label{fig:dissi}
\end{figure}

\section{Discussion and Conclusions}

A cusp is the 2D structurally unstable but topologically classifiable null point equivalent of an 
X-point. Two branches of the magnetic field lines originate in the cusp point. These two branches are 
anti-parallel and resemble the X-point anti-parallel field line merging, which is the classical
scenario of magnetic reconnection. Therefore, we investigated the relation between velocity, magnetic 
field and resistivity close to a regular magnetic cusp by analyzing non-ideal Ohm's law in 2D as analogy 
to X-point reconnection.

Our topological considerations revealed that the resistivity surface is a hyperbolic paraboloid, meaning 
that the resistivity has a saddle-point structure in the vicinity of the magnetic cusp rather than a 
maximum of the resistivity like in the case of a regular structurally stable magnetic X-point 
\citep{2012AnGeo..30.1015N}.
 
Our choice of the stagnation point flow displays an inflow-outflow scenario as is typical for 
reconnection flows. Usually, it is assumed that an anomalous resistivity, i.e., a resistivity that 
requires a monotonous relation between the current density and the resistivity, guarantees a 
typical reconnection flow. In contrast to this, our calculations imply that the resistivity
takes a saddle point configuration. The resistivity needed to guarantee regular dynamics of the plasma 
in the vicinity of such a magnetic cusp configuration can consequently not be a typical anomalous 
resistivity.

Furthermore, the energy dissipation given by $\eta j_z^2$ in our configuration has also the shape 
of a hyperbolic paraboloid including a saddle point in the vicinity of the cusp. This can be seen in 
Figure~\ref{fig:dissi}. The geometrical shape of the isocontours of the dissipation are inclined with 
respect to the geometrical structure of the cusp. Usually, the dissipation is assumed to be strongest 
either in the region where the current is highest or at the location where magnetic fields seem to 
merge, here at the cusp point. Our analysis reveals, however, that this is not the case for the 
investigated mathematical cusp magnetic field. Instead, we find that the dissipation increases with 
distance from the cusp region in downstream direction. This implies that Ohmic heating increases
in regions where magnetic flux is created and transported outwards.

\acknowledgments

We thank the anonymous referee for constructive comments that helped to improve our manuscript.
This research made use of the NASA Astrophysics Data System (ADS). D.N. acknowledges financial support 
from the Czech Science Foundation (GA\,\v{C}R, grant numbers 16-05011S). 
The Astronomical Institute Ond\v{r}ejov is supported by the project RVO:67985815.

\bibliographystyle{aasjournal}
\bibliography{ms}

\end{document}